# Experimental Observation of Strong Coupling Between an Epsilon-Near-Zero Mode in a Deep Subwavelength Nanofilm and a Gap Plasmon Mode

Joshua R. Hendrickson[1*], Shivashankar Vangala[1,2], Chandriker K. Dass[1,2], Ricky Gibson[1,3], Kevin Leedy[1], Dennis Walker[1], Justin W. Cleary[1], Ting S. Luk[4], and Junpeng Guo[5*]

[1]Air Force Research Laboratory, Sensors Directorate, Wright-Patterson AFB, OH 45433, USA
[2]KBRWyle Laboratories, Inc., Dayton, OH 45431, USA
[3]University of Dayton Research Institute, Dayton, OH 45469, USA
[4]Sandia National Labs, Albuquerque, New Mexico, 87185, USA
[5]Department of Electrical and Computer Engineering, University of Alabama in Huntsville, Huntsville, Alabama 35899, USA

**Abstract:** Strong coupling is a phenomenon which occurs when the interaction between two resonance systems is so strong that the oscillatory energy exchange between them exceeds all dissipative loss channels. Each resonance can then no longer be described individually but only as a part of the coupled, hybrid system. Here, we show that strong coupling can occur in a deep subwavelength nanofilm supporting an epsilon near zero mode which is integrated into a metal-insulator-metal gap plasmon structure. To generate an epsilon near zero mode resonance in the short-wave infrared region, an indium tin oxide nanofilm of ~$\lambda$/100 thickness is used. A polariton splitting value of 27%, corresponding to a normalized coupling rate of 0.135, is experimentally demonstrated. Simulations indicate that much larger coupling rates, well within the ultra-strong regime where the energy exchange rate is comparable with the frequency of light, are possible.

A number of interesting and diverse strongly coupled [1] systems, such as single atoms in cavities [2], quantum dots in photonic crystals [3], plasmonic resonators coupled with cyclotron resonances [4], as well as others [5-10], have been studied in the past. Ultra-strong coupling, on the other hand, has been a bit more elusive [11-15]. It is generally accepted that one has entered this regime once the ratio of the Rabi Frequency, $\Omega_R$, to the bare resonance frequency, $\omega_o$, exceeds 0.1. This implies that the energy scale of the interaction strength is on-par with all other energy scales of the system. In these coupling regimes, the dispersion curves avoid crossing each other, lifting the would-be degeneracy if they were weakly coupled. The magnitude of the splitting between the two modes is directly proportional to the coupling strength. To maximize the coupling, one needs to ensure there is sufficient spatial and spectral overlap of the two resonances as well as alignment of the electric field directions, or, in the case of coupling with optical emitters, alignment of the electric field and the emitter's dipole orientation. In this work, we investigate the strong coupling phenomena of epsilon near zero (ENZ) modes to gap plasmon modes in a structure composed of a subwavelength nanofilm integrated into a metal-insulator-metal based plasmonic patch antenna. At zero detuning the normalized coupling strength is shown to exceed the ultra-strong limit.

In order to generate a gap plasmon mode, a patch antenna design was utilized. The patch antenna, first proposed over half a century ago and commonly used in 1-6 GHz applications, is essentially composed of rectangular or circular metal patches on top of a metallic ground plane separated by a dielectric spacer layer [16]. Due to advancements in nanofabrication technologies, their operational wavelength has now been extended to the optical regime, through significant reduction of the layer thicknesses and patch sizes. Near complete absorption of light has been demonstrated in such structures due to the excitation of a gap plasmon resonance [17]. By multiplexing different patch sizes, both wideband [18] and multispectral [19] light absorption has also been achieved. When incident optical radiation is resonant with the gap plasmon mode, a strong electromagnetic field becomes confined in the dielectric-gap region between the patterned optical antenna patches and the metallic ground plane. The electric field enhancement of this gap plasmon mode is in the normal direction of the dielectric layer and, therefore, is an excellent candidate for coupling to ENZ modes in subwavelength nanofilms which also have an out-of-plane electric field component.

Epsilon near zero modes are strong localized resonance modes that only exist in deep-subwavelength material nanostructures at wavelengths where the electric permittivity is close to zero [20-22]. It is known that ENZ modes in nanofilms cannot be excited at normal incidence since the ENZ mode is oriented out-of-plane with respect to the nanofilm [23]. However, this attribute ensures that the field orientation of the ENZ mode and gap plasmon mode will be matched. As for the spatial overlap requirement for strong coupling, since ENZ modes are supported in nm-scale film thicknesses they can easily be integrated directly into a dielectric spacer region. Furthermore, ENZ modes have a very large density of states which enhances the light matter coupling strength [24]. Owing to the nature of the patch antennae design, even though both the gap plasmon mode and the ENZ nanofilm mode are oriented normal to the surface, the entire system can be excited and characterized via normal incidence illumination.

For an ENZ mode to exist at least two conditions must be met: the film material must have a real part of the dielectric constant which is equal, or very near, zero at the wavelength of interest and the film thickness must be on the order of the plasma wavelength divided by 50 [25]. ENZ materials which can potentially support ENZ modes can either be found in nature, for example near the bulk plasma frequency of metals, at strong phonon excitation wavelengths in dielectrics, in quantum well structures, and so forth, or they can be artificially engineered through metamaterial/metasurface design [26-28]. Of particular interest are the doped oxides like indium tin oxide (ITO), cadmium oxide, and aluminum or gallium doped zinc oxide [29- 31]. These materials can have a doping tunable ENZ wavelength in the near to long-wave infrared regions and, through carrier injection/depletion, can serve as actively tunable optoelectronic devices [32, 33]. Applications for ENZ materials include engineering of electromagnetic wave-fronts [34], broadband and tunable absorbers [35, 36], isolators [37], enhanced nonlinear response [38, 39], and myriad other proposed areas [40- 42]. In our work, we choose a subwavelength thick ITO nanofilm that has an ENZ wavelength in the short-wave infrared at its plasma frequency.

**Experimental Results:**
The gap-plasmon resonance structure under consideration is shown schematically in Figure 1(a). It consists of a periodically patterned array of metal squares, a subwavelength nanofilm supporting an ENZ mode, a dielectric spacer layer, and an optically thick metallic ground plane. By changing the width of the gold squares, the gap plasmon resonance can by swept through the ENZ resonance allowing us to deterministically adjust the detuning between the two resonant modes. The obtained spectra is then compared to a reference sample which only supports a gap-plasmon resonance; all other parameters remained identical with the exception that the reference sample does not contain an ENZ nanofilm.

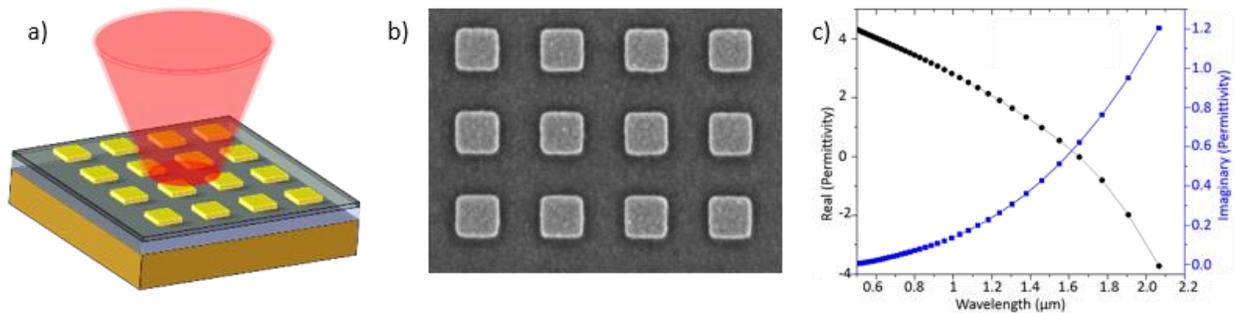

Figure 1 (a) Schematic of device consisting of a patterned periodic array of gold squares, an ENZ nanofilm, a dielectric spacer layer, and an optically thick metallic film on a substrate; (b) top down SEM image of patterned gold squares on ITO nanofilm; (c) permittivity values of a 12nm ITO film obtained from ellipsometry measurements

Preliminary studies of ITO thin films were performed in order to determine the ENZ wavelength under a given set of deposition conditions. Using pulsed laser deposition, a 12 nm thick film of ITO was deposited onto a substrate consisting of a 100 nm $SiO_2$ layer on top of a high resistivity Si wafer. Ellipsometry measurements were then performed to determine the electric permittivity values of the ITO film. As can be seen in Figure 1c), the real part of the permittivity has a zero crossing at approximately 1.65 μm. Note that the ENZ wavelength for ITO can be tuned to other values by changing the doping concentration and/or growth parameters.

For the short-wave infrared devices, an optically thick titanium (10%) / tungsten (90%) ground plane was deposited by electron beam evaporation onto a silicon substrate, followed by evaporation of a 40 nm SiO$_2$ film for device A and a 140 nm SiO$_2$ film for device C. A 12 nm and a 20 nm ITO film were then deposited on devices A and C, respectively. Finally, 16 separate gold square arrays were fabricated by using a standard electron beam lithography, evaporation, and lift-off process. Each array was 50 μm × 50 μm in size and consisted of 60 nm thick gold squares with 900 nm periodicity. The size of the square width was adjusted from one array to the next, spanning 180 nm – 480 nm in 20 nm steps. A top down SEM image of a representative device is shown in Figure 1(b). A second set of reference samples, B and D, were also fabricated where the ITO film was not deposited. A summary of the layer structures for the various devices is given in table 1.

Table 1 Structure parameters of fabricated devices

| Sample ID | SiO$_2$ spacer layer thickness (nm) | ITO layer thickness (nm) |
|---|---|---|
| A (ITO-1) | 40 | 12 |
| B (ITO -1R) | 40 | 0 |
| C (ITO-2) | 140 | 20 |
| D (ITO-2R) | 140 | 0 |

All devices were characterized by using a microscope coupled FTIR in reflection geometry and all reflection data was normalized to that of a high quality gold mirror. Figure 2a) shows the reflectivity curves from device B, the reference sample with a 40 nm SiO$_2$ spacer layer and no ENZ ITO layer. For this device the gap plasmon resonance can be seen to linearly shift from 900 nm to 2500 nm as the square size increases from 180 nm to 480 nm. In comparison, Figure 2(b) shows the reflectivity spectra for device A which is identical to B with the exception that a 12nm ITO film is now located between the SiO$_2$ spacer layer and the patterned metal squares. The spectra for the coupled device is markedly different than that of the reference sample. As the plasmon resonance is again shifted by adjusting the square width over the same 180 nm to 480 nm range, a clear anti-crossing behavior is observed, verifying strong polariton coupling between the gap plasmon mode and the ENZ mode. This is further represented in Figure 2(c) where the wavelengths of the reflectivity dips for both samples are plotted as a function of square width. In this dispersion curve, the coupled system resonances repel one another and a minimum splitting of 2$\Omega_R$ = 450 nm is observed for a square width of 300 nm, where both peaks of the strongly coupled polariton modes are symmetrically positioned around a wavelength of 1.65 μm; i.e., at zero detuning between the bare gap plasmon mode and the bare ENZ mode resonances. This corresponds to a Rabi splitting of 27% and a normalized coupling rate of ($\Omega_R/\omega_o$) = (225nm/1650nm) = 0.136, placing the coupled system firmly into the ultra-strong regime.

The dispersion curve for the reference device appears slightly blue-shifted as compared to the gap plasmon like resonance in the coupled device, at both large positive and large negative detuning. This is not unexpected as the ITO film changes the effective refractive index of the spacer layer region. In other words, at large detuning, where coupling is minimal, the gap plasmon mode of the ITO containing device will not exactly match that of the reference device because the added ITO film slightly changes the spacer layer effective index. In fact, it is because of this reason that we chose the specific layer structure for the experimental studies. Simulations, which are presented in a later section, show that even stronger coupling can be obtained for thinner SiO$_2$ layers. However, the comparison with a reference sample is no longer as easy to interpret, due to both the ITO effect on the effective index, as well as the reduced depth of the reflectivity dip for the plasmon mode which is directly related to the thickness of the spacer layer.

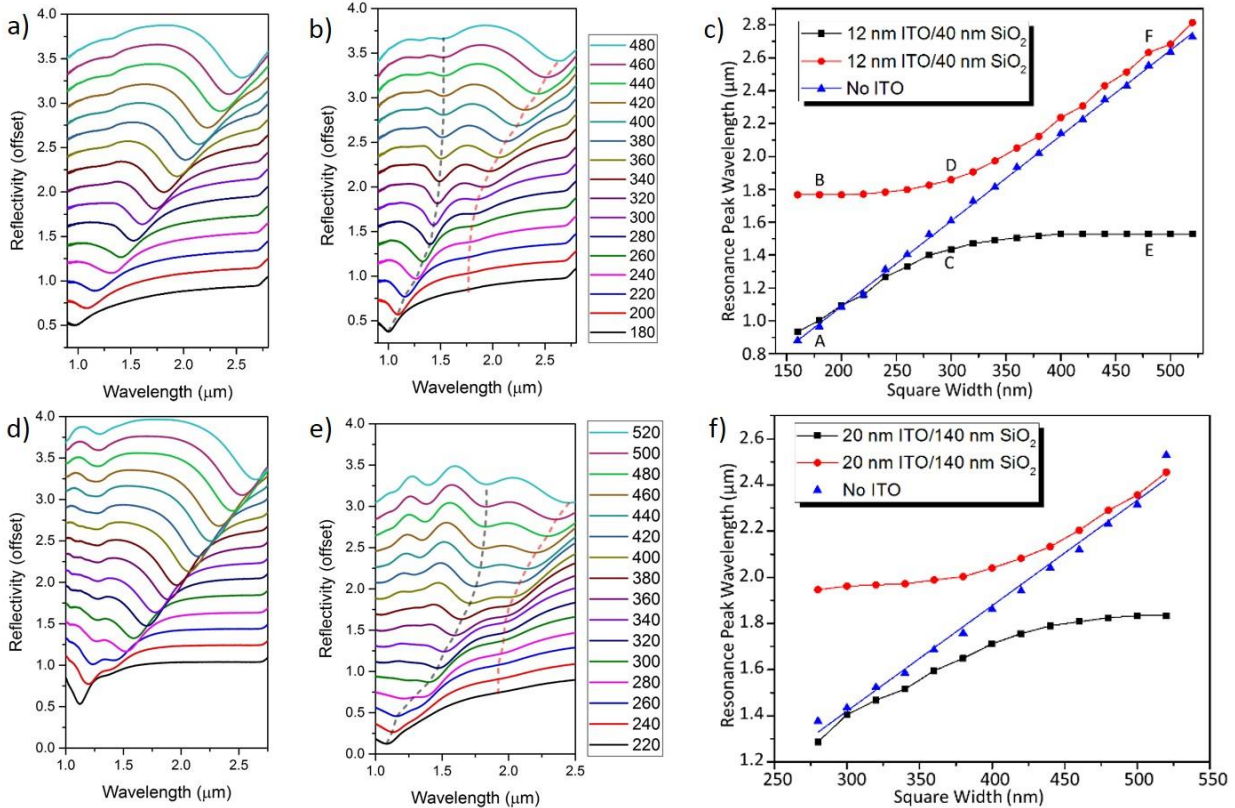

Figure 2 Measured reflectivity spectra for the 40 nm SiO$_2$ samples without (a) and with (b) a 12 nm ITO film and the 140 nm SiO$_2$ samples without (d) and with (e) a 20 nm ITO film. For the ITO devices, dashed lines are fitted to the upper and lower polariton branches. Anti-crossing curve for the 12 nm ITO (c) and 20 nm ITO (f) coupled devices. Red circles and black squares correspond to the upper and lower polariton branches, respectively. Blue triangles correspond to the bare gap plasmon resonance. The A-F labelling in (c) refers to locations where electric and magnetic field profiles were taken.

The experimental reflectivity spectra for the second set of devices containing a 140 nm SiO$_2$ layer both with and without a 20 nm ITO layer, along with the corresponding anti-crossing curve, is also displayed in Figure 2. Again, strong coupling is observed, this time with a minimum Rabi splitting of 325 nm. At first, one might have expected a larger Rabi splitting for the thicker ITO layer, as described elsewhere [20], however, the reason we observe a reduced splitting here is due to the additional thickness of the spacer layer resulting in a larger mode volume for the gap plasmon resonance. This reduces the mode volume overlap of the gap plasmon mode with the ENZ mode, resulting in weaker interaction strength and, therefore, reduced coupling strength. Another difference with this set of devices are the additional peaks that arise on the short wavelength side as the patterned gold square width is increased, as well as the redshift of the zero detuning wavelength. The additional modes are simply related to the gap plasmon resonances. As for the redshift of the ENZ wavelength, this can be attributed to unintentional, slight changes in the deposition process of the ITO nanofilm.

**Numerical Simulations:**
Simulations of devices were performed using the software package FDTD by Lumerical Inc. [43]. Permittivity values for the Ti-W ground plane and the ITO ENZ layer were imported from ellipsometry measurements while database values were used for SiO$_2$ and gold. Figure 4 shows surface plots of the simulated reflectivity for the same layer structure as devices A and B (40 nm SiO$_2$ with and without 12 nm ITO). These results match up very well with the experimental data and anti-crossing is clearly evident in the simulated 12 nm ITO device. In the coupled device, for large detuning where the square is small, only the gap plasmon resonance reflectivity dip can be seen and not the ENZ mode resonance. This is due to the fact that the ENZ mode cannot be directly accessed with normal incidence excitation. In such an excitation scheme the ENZ mode, therefore, needs to be coupled to the gap plasmon mode in order to be excited. When the detuning is large there is essentially no coupling and, hence, the ENZ mode is not observable. The simulations do show ENZ mode excitation at large positive detuning and this is merely a consequence of the bandwidth of the gap plasmon mode for larger square widths. In the simulated spectra, one can see that the bandwidth of the bare gap plasmon resonance is narrower for smaller square widths and wider for larger square widths. This results in increased residual coupling for positive detuning values as compared to their equivalent negative detuning values.

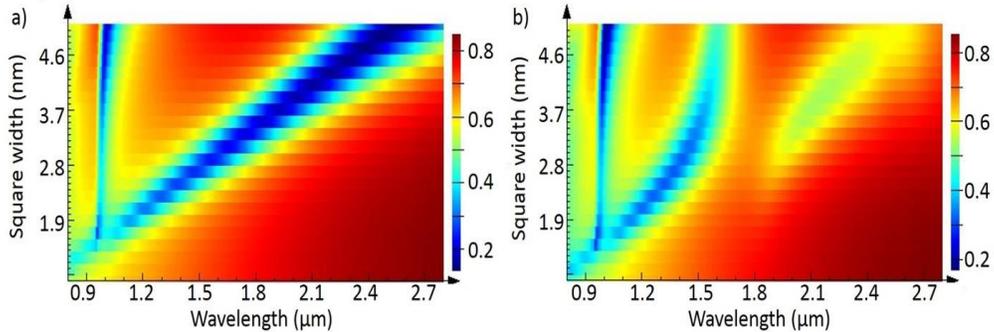

Figure 3 Simulated reflectivity as a function of wavelength and square width for a 40 nm SiO2 layer device without (a) and with (b) a 12 nm ITO layer.

It is well known that in a strongly coupled system, at zero detuning the two split modes cannot be viewed as two independent and separate modes; i.e., in the specific case studied in this work, at zero detuning there is no longer a gap plasmon mode and an ENZ mode but only coupled hybrid modes. As such, the field profiles of the two hybrid resonances at zero detuning should be very similar. In the large detuning cases, where the coupling is greatly reduced, the independent nature of each mode is recovered and one resonance should display a gap plasmon mode profile and the other an ENZ mode profile. Indeed, this is the case, as evidenced by the simulated electric and magnetic field magnitude plots displayed in Figure 4, for a device with the layer structure of A (40 nm $SiO_2$ and 12 nm ITO). The field profiles for each of the two polariton modes were obtained for arrays with square widths of 180 nm, 300 nm, and 480 nm, corresponding to large negative detuning, zero detuning, and large positive detuning, respectively. For clarity, the exact positions where the field profiles were taken is labelled in the anti-crossing curve in Figure 2. At large negative detuning the electric field of the short wavelength polariton resonance permeates throughout the spacer layer region, as expected for a gap plasmon resonance, while the long wavelength resonance is highly confined to the ITO nanofilm, as expected for an ENZ mode resonance. For large positive detuning the roles become reversed and the short wavelength polariton resonance now displays an ENZ mode profile while the long wavelength resonance displays a gap plasmon mode profile. More interestingly, when the detuning is zero, the two hybrid resonances display nearly identical field profiles. The magnetic field profiles also support the same conclusion. At large detuning the gap-plasmon-like polariton mode shows a strong magnetic resonance directly under the metal square while the ENZ-like polariton mode shows a very weak magnetic field profile. At zero detuning the magnetic field profiles are very similar. These forms of the electric and magnetic field profiles are a direct result of the strong coupling between the gap plasmon mode and the ENZ mode.

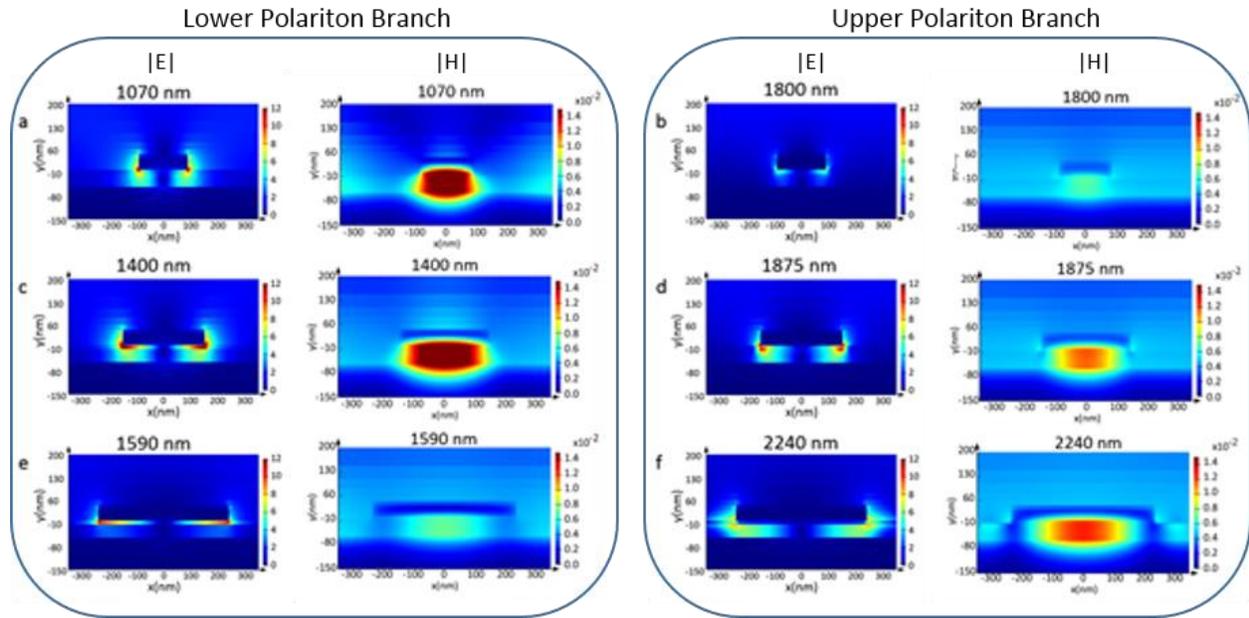

**Figure 4** Electric and magnetic field magnitude of the 12 nm ITO device for large negative detuning (a-b), zero detuning (c-d), and large positive detuning (e-f). (a), (c), and (e) correspond to the shorter wavelength polariton mode and (b), (d), and (f) correspond to the longer wavelength polariton mode. In figure 2 each of these wavelength positions is pinpointed on the anti-crossing curve.

In order to map out the parameter space of possible coupling strengths, additional simulations were carried out for 12 nm, 20 nm, and 30 nm ITO nanofilms, as a function of both $SiO_2$ spacer layer thickness and square width. From these simulations, the Rabi splitting values at zero detuning were recorded and plotted as a function of $SiO_2$ thickness. These results are presented in Figure 5. As the $SiO_2$ thickness is reduced, the polariton splitting increases. This increased coupling strength is a result of the increased spatial mode overlap for thinner $SiO_2$ layers versus thicker $SiO_2$ layers. However, one should keep in mind that the magnitude of the bare gap plasmon resonance dip is directly related to the spacer layer thickness, therefore, while the coupling strength is larger for thinner $SiO_2$ layers, the reflectivity dips (absorption peaks) may not be as pronounced. Another observation from the coupling strength investigation is that thicker ITO nanofilms show increased splitting over thinner nanofilms, as expected from previous studies of ENZ modes [20]. These simulations show that with proper device design Rabi splitting of over 50% is possible in our device configuration.

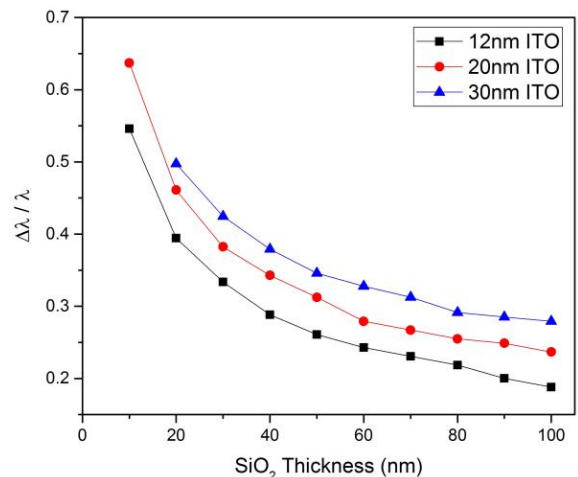

**Figure 5** Polariton splitting at zero detuning for 12 nm, 20 nm, and 30 nm ITO films as a function of $SiO_2$ spacer layer thickness.

**Conclusion:**

In conclusion, strong coupling between a gap plasmon mode resonance and an ENZ mode resonance has been experimentally demonstrated. In the short-wavelength infrared spectral region, the ENZ mode was excited in a 12 nm ITO nanofilm integrated into a metal-insulator-metal gap plasmon structure. Experimental results show a Rabi splitting value of 27% while simulations show that values beyond 50% are achievable. Such strong polariton splitting, along with the potential to actively tune ENZ modes through carrier injection/depletion, opens up the door to a number of new devices and applications. One could also envision replacing the ENZ nanofilm with other materials containing out-of-plane oriented electric fields or dipoles, such as dark excitons in 2D materials.

**Acknowledgments:**

JH and JC acknowledge support from the Air Force Office of Scientific Research (Program Manager Dr. Gernot Pomrenke) under contract number FA9550-15RYCOR159 and FA9550-15RYCOR162, respectively. JG acknowledges support by the National Science Foundation under the award no. 1158862. This work was supported, in part, by the Center for Integrated Nanotechnologies, an Office of Science User Facility operated for the US DOE Office of Science.

**Author Contributions:**

J.R.H, T.S.L., and J.G. conceived of the project idea. S.V. led all aspects of device fabrication. J.W.C. performed ellipsometry measurements. K.L. performed PLD of the ITO films. D.W. performed electron beam lithography. J.R.H and C.K.D simulated the devices. J.R.H. and R.G. experimentally characterized the devices. J.R.H and J.G wrote, and all authors edited, the manuscript. J.R.H and J.G. supervised the project.

**Methods:**

ITO Deposition:
ITO films were deposited in a Neocera Pioneer 180 pulsed laser deposition system with a KrF excimer laser (Coherent COMPex Pro 110, λ=248 nm, 10ns pulse duration). The laser operates at a repetition rate of 30 Hz and an energy density of 3 J/cm$^2$, producing a deposition rate of 22 nm/min. The deposition temperature was 300° C and pressure was 1.3 Pa in a mixture of 5% $O_2$ and 95% Ar. The target was a 50 mm diameter by 6 mm thick sintered 90 % In2O3 / 10% SnO2 ceramic disk (99.99% purity).

Ellipsometry:
J.A Woollam VUV-VASE Near infrared ellipsometry was used to characterize the permittivity of materials. Metallic films characterized were determined to be optically thick and thus are suitable for ground planes. All transparent or ENZ materials used Kramers-Kronig consistent models to determine optical constants in the wavelength range of interest.

Simulations:
All simulations were performed using Lumerical FDTD. Permittivity values for gold (Palik) and SiO2 were taken from the standard Lumerical materials database. The permittivity of TiW and ITO were extracted from ellipsometry measurements. For a given device, a single unit cell was simulated using periodic boundary conditions. An override mesh of 2nm in the direction normal to the film, and 5nm in-plane, was used around the dielectric spacer, ENZ nanofilm, and gold square layers.

Characterization:
Reflectivity spectra was obtained using a Bruker Vertex 80V with a coupled Hyperion microscope. A 15×, 0.4 NA reflecting objective was used. All spectra were normalized to that of a high quality gold film.